\documentclass[10pt,twoside]{amsart}
\usepackage{amssymb}
\usepackage{latexsym}
\usepackage{fullpage}
\usepackage{a4,latexsym,rotating,amsmath, amsfonts}
\usepackage{color}
\usepackage{graphicx}

\usepackage{amscd}
\usepackage{oldgerm}

\def\N{\mathbb{N}}
\def\Z{\mathbb{Z}}
\def\R{\mathbb{R}}

\def\SSS{\mathbb{S}}
\def\vol{\rm vol}
\newcommand{\SecondFF}{\operatorname{\mathit{II}}}    

\DeclareMathOperator{\ric}{\mt{ric}}

\newcommand{\grad}{\operatorname{grad}} 
\newcommand{\tr}{\operatorname{tr}} 
\newcommand{\Ext}{\operatorname{Ext}} 
\newcommand{\cl}{\operatorname{cl}} 
\newcommand{\Ric}{\operatorname{Ric}} 

\renewcommand{\S}{\mathbb{S}} 

\def\1{\mathbf{1}}
\def\:{\lrcorner}
\def\#{\sharp}

\def\l{\lambda}
\def\a{\alpha}
\def\b{\beta}
\def\g{\gamma}
\def\d{\delta}
\def\e{\epsilon}

\def\o{\circ}

\def\x{\otimes}
\def\<#1,#2>{\langle#1,\,#2\rangle}

\def\S{\mathbb{S}\,}

\def\qed{\ensuremath{\quad\Box\quad}}
\def\pfill{\par\Vskip2mm plus1mm minus1mm\noindent}
\def\inv#1{\raise.1em\hbox to 0pt{$^{-1}$\hss}_{#1}\;}

\def\V{\noindent}

\newcommand{\ov}{\overline}

\newcommand{\bean}{\begin{eqnarray*}}
\newcommand{\eean}{\end{eqnarray*}}
\newcommand{\benu}{\begin{enumerate}}
\newcommand{\eenu}{\end{enumerate}}
\newcommand{\eea}{\end{eqnarray}}
\newcommand{\bea}{\begin{eqnarray}}
\newtheorem{Theorem}{Theorem}
\newtheorem{Lemma}[Theorem]{Lemma}
\newtheorem{Corollary}[Theorem]{Corollary}
\newtheorem{Definition}[Theorem]{Definition}
\newtheorem{Example}[Theorem]{Example}
\newtheorem{Remark}[Theorem]{Remark}
\newtheorem{Proposition}[Theorem]{Proposition}

\def\BH{{\rm BH}}
\def\ext{{\rm ext}}

\def\Ein{{\rm Ein}}
\def\ric{{\rm ric}}
\def\exp{{\rm exp}}

\def\1{\mathbf{1}}
\def\:{\lrcorner}
\def\#{\sharp}

\def\x{\otimes}
\def\qed{\ensuremath{\quad\Box\quad}}
\def\pfill{\par\Vskip2mm plus1mm minus1mm\noindent}
\def\inv#1{\raise.1em\hbox to 0pt{$^{-1}$\hss}_{#1}\;}

\def\V{\noindent}

\newcommand{\be}{\begin{equation}}
\newcommand{\ee}{\end{equation}}

\newcommand{\ben}{\begin{enumerate}}
\newcommand{\een}{\end{enumerate}}
\newcommand{\bit}{\begin{itemize}}
\newcommand{\eit}{\end{itemize}}
\newcommand{\edoc}{\end{document}}

\newcommand{\bdefi}{\begin{Definition}}
\newcommand{\btheo}{\begin{Theorem}}
\newcommand{\bprop}{\begin{Proposition}}
\newcommand{\brema}{\begin{Remark}}
\newcommand{\bcoro}{\begin{Corollary}}
\newcommand{\blemm}{\begin{Lemma}}
\newcommand{\bexam}{\begin{Example}}

\newcommand{\edefi}{\end{Definition}}
\newcommand{\etheo}{\end{Theorem}}
\newcommand{\eprop}{\end{Proposition}}
\newcommand{\erema}{\end{Remark}}
\newcommand{\ecoro}{\end{Corollary}}
\newcommand{\elemm}{\end{Lemma}}
\newcommand{\eexam}{\end{Example}}

\title{Black holes in Einstein-Maxwell Theory}

\author{Olaf M\"uller}

\begin{document}

\begin{abstract}
\V We prove variants of known singularity theorems ensuring the existence of a region of finite lifetime that are particularly well applicable  if the solution admits a conformal extension, a property satisfied e.g. by maximal Cauchy developments of Einstein-Maxwell initial values close to the trivial ones.
\end{abstract}

\maketitle

\section{Introduction}

\V A fundamental question of mathematical relativity is which initial values evolve to singularities of some kind, in particular to black holes. There is a standard definition of 'black hole' for asymptotically flat spacetimes, generalized in \cite{oM} to arbitrary spacetimes. According to that latter definition, a point in $(M,g)$ is called {\em black} iff there is no future timelike curve of infinite length starting at $p$, and the {\em black hole ${\rm BH} (M,g)$ of $(M,g)$} is the subset of all black points. Of course, ${\rm BH} (M,g)$ can be empty. In the one-ended asymptotically flat case, this notion coincides with the definition $ {\rm BH}_M := M \setminus I^-(\mathcal{J}_{{\rm max}}^+)$ where future null infinity $\mathcal{J}^+_{{\rm max}}$ is not an additional datum as usual but simply the set of {\em all} ideal end points of null geodesics of infinite affine parameter, see Section \ref{Hoop}. Moreover, this notion of 'black hole' is exactly the one appearing implicitly in the conclusion of Hawking's singularity theorem, as explained below. The primary application we will have in mind is Einstein-Maxwell theory. Maxwell theory is considered to be a relatively fundamental theory (one building block of the Standard Model), which can be assumed to leave its domain of validity at a much larger energy scale than effective models like dust or other fluid equations. This is of some importance as due to the definition of blackness we must have the ambition to include the high-energetic interior of the black hole in our model. Moreover, Einstein theory (gravity) and Maxwell theory (electromagnetism) are {\em the only} long-range fundamental interactions that are well-tested as classical field theories, and their canonical coupling, Einstein-Maxwell theory --- considered in greater detail in Section \ref{Compactifications} --- has also been shown to be relevant as a classical field theory, e.g. in modelling the early, radiation-dominated, universe. The reasons above explain why it is considered an important issue to understand the space of solutions of Einstein-Maxwell theory, e.g. their timelike completeness. 

\bigskip

\V The starting point of our analysis are the classical singularity theorems of Hawking and Penrose. Penrose's singularity theorem \cite{rP} is one of the cornerstones of mathematical relativity. It needs three conditions: Existence of a noncompact Cauchy surface in the spacetime, the {\em null convergence condition (NCC)} $\ric (v,v) \geq 0$ for all $v \in TM$ null (implied trivially by the {\em timelike convergence condition (TCC)} $\ric (v,v) \geq 0$ for all $v \in TM$ causal, which in turn is satisfied by all solutions of many Einstein-matter models like Einstein-Maxwell Theory), and the existence of a trapped compact spacelike codimension-$2$ submanifold $N$. Here, a submanifold $N$ is called {\em trapped} iff its mean-curvature vector field is future. It is called {\em uniformly trapped} iff it is trapped and if there is $c>0$ s.t. $g(H_N, H_N) < - c^2$. Of course, any compact trapped submanifold is uniformly trapped. Under these three conditions, Penrose's theorem states that the spacetime contains an incomplete $C^0$-inextendible future lightlike geodesic starting at $N$. Still, to connect this to the usual idea of 'black hole', one could ask if there are also complete null geodesics from the trapped submanifold. Moreover,  one is more interested in timelike than in null curves (as the former describe possible trajectories of massive objects, like observers). Now, answering the question whether a trapped submanifold implies the appearance of a black hole would imply an analogue of Penrose's theorem for {\em timelike} curves instead of null curves.  There is an appropriate singularity theorem for timelike curves, namely Hawking's theorem \cite{sH}. Here, replacing the NCC by the TCC, we moreover assume the existence of a uniformly trapped {\em Cauchy surface} (which is necessarily of codimension $1$). Unfortunately, in the physically relevant spatially asymptotically flat case there is no uniformly trapped Cauchy surface, thus Hawking's theorem is not applicable to this important class of models. Therefore it would be desirable to have a theorem for the asymptotically flat case whose hypothesis is of Penrose's type and whose conclusion is of Hawking's type. 
First steps in this direction are taken by the main results of this article, accounted for in the following part of this section.

\bigskip

\V An important condition in our context is the existence of a conformal extension, this notion being a slight generalization of the one of conformal compactification. These concepts fits very well the conformal invariance of the theory we will mainly consider, which is Maxwell Theory in spacetime dimension $4$. In Section \ref{Compactifications} we first confirm folk knowledge by giving a result (Theorem \ref{X}) on nonexistence of sufficiently smooth conformal compactifications. Then we define the notion of {\em conformal $C^k$-extension} of a globally hyperbolic manifold $(M,g)$, which is an open conformal embedding $E$ into another globally hyperbolic manifold $(N,h)$ with $E$ and $h$ of regularity $C^k$ such that the closure $\cl (E(M))$ of its image is causally convex and future compact (details in Sec. \ref{Compactifications}). This notion has been introduced in \cite{nGoM} where it proved useful to show that on conformally extendible spacetimes, Dirac-Higgs-Yang-Mills (DHYM) theories have a well-posed initial value problem for small initial values\footnote{modulo gauges, and each choice of a Cauchy surface corresponds to a smooth Cauchy section of the gauge orbits.}. Note that this definition does not make any statement on vanishing or nonvanishing of the conformal factor at the boundary, nor on regularity of the latter. As usual, for a subset $A$ of a spacetime $(N,h)$  we define $ \partial^\pm A := \{ x \in \partial A \vert I^\mp(x) \cap A \neq \emptyset \} $. Let $l \in \N \setminus \{ 0 \}, s \in [0; \infty), p \in \partial^+E(M)$, then a conformal $C^k$-extension $E$ for $k \geq 1$ is called {\em strong of degree $l$ and order $s$ at $p$} if for the conformal factor $\Omega$ with $g= \Omega h $, the function $\omega^s:= \Omega^{-s}$, as a function on $E(M)$, can be extended to a $C^l$ function on $N$, such that there is a Cauchy temporal function $T$ of $(N,h)$ with $(\grad T) \omega^s  \neq 0$ in a neighborhood of $p$ (an elementary calculation shows that $s$ as above is unique at every zero locus of $\omega^s$; the default entries are $l=k$ and $s= 1/2$). Furthermore, we say that a spacetime has {\em standard spatial infinity} if and only if for every future curve $ c : I \rightarrow M$ with $ J^-(c(I)) \cap S $ noncompact for a Cauchy hypersurface $S$, all null future geodesic rays with image in in $ J^- (c(I))$ that are $C^0$-inextendible in $M$ have infinite affine length.

\bigskip

\V The first main theorem, proven  in Section \ref{Compactifications}, confirms the well-known fact that conformal extendibility is indeed a frequent property --- conformal extensions exist for the maximal Cauchy developments of every initial value contained in the open neighborhoods of the Euclidean metric $e_3$ considered in Zipser's PhD thesis \cite{nZ} where admissible initial values (henceforth called {\em Zipser-asymptotically flat}, for details see Section \ref{Compactifications}) are defined via weighted Sobolev spaces. In Section \ref{Compactifications} we show, starting from Zipser's result,  by considering induced values on hyperboloidal slices in an appropriate gauge:

\bigskip

\begin{Theorem}
\label{ExtendibilityIsAsymptotic}
There is $\e>0$ such that for every Zipser-asymptotically flat initial value $I:= ( S, g_0:= e_3 +\g_0 , K_0, A_0, \dot{A}_0)$ for Einstein-Maxwell theory with $\vert \g_0 \vert_{3+k} + \vert K_0 \vert_{2+k} + \vert F_0 \vert_{2+k}< \e $ (where $F:= dA$), the metric part $(M,g)$ of the maximal Cauchy development $(M,g,A)$ of $I$ admits a conformal $C^k$-extension. Moreover, $(M,g)$ has standard spatial infinity, and $ J^+ (S) \setminus J^+(C) $ has infinite timelike diameter for all compact $ C \subset S$. For every compact set $C \subset S$, there is $\e_C >0$ such that, if additionally $\vert \g_0 \vert_{3} + \vert k \vert_{2}  + \vert F_0 \vert_{2} < \e_C$, then the maximal Cauchy development of $I$ has a conformal extension $E$ strong (of order $1/2$) at every $p \in \partial^+ E(M) \setminus J^+(E(C))$.   
\end{Theorem}

\V For a spacetime $M$, a real interval $K$ and a curve $c: K \rightarrow M$, we write $I^\pm(c) := I^\pm (c(K))$ and $J^\pm (c) := J^\pm (c(K))$. In Section \ref{SCOTs} we show via the elliptic maximum principle:

\begin{Theorem}
\label{HorizontalCITIFs}
Let $(M,g)$ be an $n$-dimensional globally hyperbolic spacetime satisfying the timelike convergence condition. Let $c$ be a timelike curve in $M$. If $J^-(c)$ contains a compact maximal hypersurface $S$ with $\partial S \subset \partial J^-(c)$, then $c$ has finite length. 
\end{Theorem}

\V With essentially the same method we show:

\begin{Theorem}
\label{VariantHawking}
Let $ (M,g)$ be an $n$-dimensional globally spacetime satisfying the timelike convergence condition, let $p \in M$. Assume that $(J^-(J^+(p)),g)$ contains a compact maximal hypersurface $S$ with $\partial S \subset \partial J^-(J^+(p))$. Then $p$ is black. 
\end{Theorem}

\V The compactness condition on $J^-(c)$ resp. $J^-(J^+(p))$ implies that $J^-(c) $ resp. $ J^-(J^+(p))$ is spatially compact. Conversely, the condition on the existence of a compact maximal hypersurface is satisfied for a neighborhood of the trivial initial values for Einstein-Maxwell theory if $J^-(c) $ is spatially compact. In fact, the results by Bieri and Zipser ensure the existence of an entire foliation of $M$ by maximal hypersurfaces, thus any slice can be intersected with $J^-(c)$ to produce $S$ as required. 
A second source of examples admitting such maximal hypersurfaces consists in the appropriate, Schwarzschild-like, asymptotic behaviour of the metric along the singularity, as we will see in the same section.

\V There is third possibility, presented in Section 3, to construct a compact maximal hypersurface $S$ in $I^-(c)$ with $\partial S \subset \partial I^-(c)$ under the condition of a conformal compactification via the mean curvature flow. To explain the result, we define the conformal property of having a spacelike endpoint: For a $C^0$-inextendible future curve $c:A \rightarrow M $, where $A$ is a real interval, we write $I^-(c):= I^-(c(A))$ and $J^-(c) := J^-(c(A))$, and $c$ is said to {\em have a spacelike endpoint} iff $I^- (c) \setminus I^+(c(t)) $ is spatially precompact for every $t \in A$. A point $p \in M$ is called {\em spatially topped} iff every $C^0$-inextendible timelike future curve $c:[0, \infty) $ from $p$ has a spacelike endpoint. This property implies spatial compactness of $J^-(J^+(p))$ but it is a bit stronger than the latter property \footnote{In the terminology of \cite{oM}, spatial toppedness is equivalent to the property of points to be lower-shielded, i.e. to be the starting point only of non-dominating CITIFs.}. Then we prove in Section 3:

\begin{Theorem}
\label{EmergingMaximal}
Let $(M,g)$ be an $n$-dimensional spaetime. Let $c$ be a timelike curve with a spatial endpoint and let $I^-(c)$ have a conformal extension $E: I^-(c) \rightarrow N$ of order $ s> \frac{n-1}{4}$. Then $I^-(c)$ contains a maximal hypersurface $S$ with $\partial S \subset \partial I^-(c)$. 
\end{Theorem}

\V An immediate consequence is that a spatially topped point $p$ is black, if the causal boundary of $I^+(p)$ is 'pointwise conformally extendable' in the sense that for every $C^0$-inextendable future curve $c$ from $p$ we have that $I^-(c)$ admits a conformal extension. One should keep in mind that the condition on the order of the conformal extension is satisfied only for $n=3$ by Cauchy developments of initial values close to trivial ones (which admit conformal extensions of order $1/2$). However, the result could be valuable for initial values far from the trivial ones, and the method of the proof seems interesting in itself.

\V Recall that the compactness condition on $S$ ensures that $J^-(J^+(p))$ is spatially compact. As spatial toppedness and spatial compactness of $J^-(J^+(p))$ are purely conformal notions, whereas blackness of points depends on the full geometry, these theorems provide an interesting connection between the conformal structure of a spacetime and its full geometry, and the bridge between those two levels is unsurprisingly given by a condition on the Ricci tensor.

\V The reader should be aware of the fact that this statement differs substantially from the only superficially similar statement of Prop. 9.2.1 in the book of Ellis and Hawking \cite{HE}. The differences are explained in detail in Section \ref{Hoop}.

\V In Section \ref{jacobi}, we then get a slight generalization of Penrose's singularity theorem, Theorem \ref{VariantPenrose}, replacing trapped surfaces by outer trapped surfaces.

\V Implications for the electromagnetic potential $A$ in the situation as in the conclusion of Theorem \ref{ExtendibilityIsAsymptotic} are considered immediately after the proof of Theorem \ref{ExtendibilityIsAsymptotic}, which is given in Section \ref{Compactifications}.

\V A further consideration in Section 5 shows that there actually are examples of spatially asymptotically flat globally hyperbolic spacetimes that satisfy the timelike convergence condition, contain an OTS and are conformally extendable. 

\bigskip

\V In a subsequent article in preparation, the abundance of outer trapped surfaces in the relevant space of initial values will be shown via other tools. 

\bigskip  
  
\V In Section \ref{discussion}, we briefly discuss possible physical interpretations of the results.

\section{Conformal compactifications and conformal extensions}
\label{Compactifications}

\V The concept of conformal compactifications has proved to be useful in Lorentzian geometry just as in Riemannian geometry. Let us recall first the basic definitions. Let $k \in \N \setminus \{0\}$. A subset $C$ of a spacetime $P$ is called {\em causally convex} iff any causal curve intersects $C$ in the image of a (possibly empty) interval, i.e. no causal curve can leave and re-enter $C$. A {\em conformal compactification} of a globally hyperbolic manifold $(M,g)$ is an open conformal embedding $E$ into another globally hyperbolic manifold $(N,h)$ with $E$ and $h$ of regularity $C^k$ such that the closure $\cl (E(M))$ of its image is causally convex and compact.

\V As every conformal compactification induces a conformal compactification of the Cauchy surfaces, our starting point is the following easy statement on possible conformal compactifications of asymptotically flat {\em Riemannian} manifolds. It is probably well-known to experts on the field, but I could not find a written statement in the literature.

\begin{Theorem}
\label{ItIsOnePoint}
Let $(S,g_0,0)$ converge to standard initial values to order $0$, i.e. \\ $\Vert (g_0)_{ij} - \delta_{ij}\Vert < c/r $ for an appropriate constant $c>0$, where $r$ is the Euclidean radius. Then every open conformal embedding of $(S,g_0)$ with precompact image is a map into a sphere the complement of whose image is exactly one point.  
\end{Theorem}

\V {\em Proof.} We show that, for every Riemannian manifold $(N,h)$ and every open conformal map $Z: M \rightarrow N$ with $h= u \cdot Z_* g_0$ on $Z(M)$ whose image is precompact, the topological boundary of $Z(S)$ consists of one point only. To that purpose, let $x_1, x_2 \in \partial Z(M)$ be given. We want to show that their $h$-geodesic distance vanishes. Consider sequences $y^i_n \in M$ converging to $x_i$, for $i \in \{ 1,2 \}$. We choose accumulation points $s_i $ of the projections $p^i_n$ of $ y_i^n$ on the $\SSS^{n-1}$ component (w.r.t. the diffeomorphism $ P: \R^n \setminus \{ 0 \} \rightarrow \R^+ \times \SSS^{n-1}  $ induced by $r$). Then the curve $k_i$ defined by $k_i(t) := P^{-1}(s_i, t)$ has finite length, and an easy triangle inequality argument shows that $k_i$ approaches $x_i$. Now let $\e >0$ and choose points $y_i$ in the image of $k_i$ with $d(x_1, y_1) < \e/6$ and $d(x_2, y_2) < \e/6$. What remains to be shown is that there is a curve of length $< \e/3$ from $y_1 $ to $y_2$. To that aim, choose an arc-length parametrized geodesic curve $\gamma$ of length $\leq 2 \pi$ in $\SSS^{n-1}$ between $s_1$ and $s_2$. Now, the curves $k_s $ (for simplicity all parametrized by $g$-arc length) are of finite $h$-length, thus for all $r_0$ there is an $r >r_0$ such that $u(k_s(r)) < \frac{\e}{12 \pi}$. Also for each $s$ there is an $r_1 (s)>0$ such that the $h$-length of $ c_s \vert_{[r_1, \infty)} $ is smaller than $\frac{\e}{6}$. By compactness, there is a finite maximum $m$ of $r_1 $ along the sphere. Now we want to construct a continuous and piecewise $C^1$ curve $\hat{\gamma}$ of length smaller than $\e/3$ projecting to $\gamma$. The strategy is to begin at a point $P^{-1} ( v, \gamma (0) ) $ for some $v>m$ and then to use the open neighborhoods $U_l$ in which $u(k_s(r)) < \e/r^2$ to progress along the sphere (i.e., with constant first component) up to the boundary of $ U_l $ and then continue in outward radial direction (i.e., with constant second component) until reaching the next neighborhood $U_{l+1}$. An easy connectedness argument ensures that using this procedure, one can cover the whole path $\gamma$. The lengths of the spherical parts of $\hat{\gamma}$ add up to at most $ \e/6$ as do the radial parts. \hfill \qed           

\bigskip

\V Unfortunately, in many physical situations, energy conditions obstruct the existence of conformal compactifications:

\begin{Theorem}
\label{X}
Let $(S, g_0, K_0, B_0)$ be asymptotically flat initial values for Einstein-matter equations obeying the dominant energy condition, $S$ spin or of dimension $\leq 7$ and with not identically vanishing matter fields $B_0$, then any open conformal embedding with precompact image is not $C^2$ at infinity.  
\end{Theorem}

\V {\em Proof.} First, theorem \ref{ItIsOnePoint} implies that any compactification would be by one point $i_0$ only. Next, the appropriate spacetime positive mass theorem \cite{eW}, \cite{PT}, \cite{EHLS} implies that the ADM mass does not vanish. Now we examine the quantities $ W \in C^\infty (M) $ defined by $ w(x):= \sum_{abcd}{\rm Riem}^{abcd} (x){\rm Riem}_{abcd} (x) =  \sum_{abcd}{\rm Weyl}^{abcd} (x){\rm Weyl}_{abcd} (x)$ and $W \in C^\infty (\R^+, \R)$, $W (r):= \int_{\partial B(p,r)} w /{{\rm vol}_{n-1} \partial B(p,r)}$, and slightly adapt (from $w$ to $W$) the example of Ashtekar-Hansen \cite{AH} showing that, as in a standard Schwarzschild slice we have $w \in \frac{M}{r^6} + O(r^{-7})$, in asymptotically flat manifolds we still have $W \in \frac{M}{r^6} + O(r^{-7})$ and for $ g' = a^2 g $ with $ a \in \frac{1}{r^2} + O(r^{-3}) $ we get $W' \in r^2 + O(r) $, thus $W'$ diverges towards $i_0$. Alternatively, reduce the general case to the Schwarzschild calculation in \cite{AH} by employing Lemma 2 of \cite{fS} (which in turn refers for some proofs to \cite{SY0} and to \cite{hB}), generalized to appropriately defined {\em almost} conformally flat manifolds (referring to blown-up Riemannian normal neighborhoods of $i_0$) via a $3 \e$-argument involving the tensors $K_0, B_0$. \hfill \qed 

\bigskip

\V This implies that one has to generalize the notion of conformal compactification if one wants it to be at least $C^2$. Our choice is the one of conformal extension. A subset of a globally hyperbolic manifold is called {\em future compact} iff its intersection with every future set is compact, or equivalently, iff it is contained in the past of a compact set. As defined in the introduction, a conformal extension is an open conformal embedding $E: (M,g) \rightarrow (N,h) $ between two globally hyperbolic manifolds such that ${\rm cl} (E(M))$ is causally convex and future compact. Thus the notion of conformal extensions arises from the one of conformal compactifications by just replacing the requirement of compactness with the one of future compactness. It has been shown in \cite{nGoM} that conformal extensions can be very useful for PDE questions, e.g. it leads to well-posedness of the small-initial value problem in Dirac-Higgs-Yang-Mills theories. The usual way around the problem at spatial infinity shown above is to just renounce higher regularity of the metric of $(N,h)$ at $i_0$. If in such a generalized conformal compactification $E: M \rightarrow N$, weaker differentiability at $i_0$ is permitted then we can construct a conformal extension in the sense above by simply replacing $N$ with $  I^+(E(M))$, which is a future set in $N$ and thus globally hyperbolic. In this sense the notion of conformal extension is more generally applicable, still offering the same analytical benefit. Now first of all, it is easy to see that conformal exensions satisfy a uniqueness property:

\begin{Theorem}
\label{UniquenessConfExt}
Let $(M,g)$ and $(N,h)$ be given. A conformal extension of $(M,g)$ into $(N,h)$ is uniquely given by its values on a $g$-Cauchy set $S$.
\end{Theorem}

\V {\em Proof.} The core step of the proof is that any point $p$ in $J^+(S)$ is determined uniquely by $J^-(p) \cap S$, a datum which is independent of the choice of the conformal extension, if its value on $S$ is given. This can be proven along the lines of the following lemma (a variant of Theorem 3 in \cite{oM}):

\begin{Lemma}
\label{catcher}

Given $p,q \in M$, $p \notin J^-(q)$, then there is a $C^0$-inextendible past timelike past curve $c$ from $p$ not intersecting $J^-(q)$.

\end{Lemma}

\V {\bf Proof of the lemma:} Let $q^{+} \in I^+(q)$ and consider, for a Cauchy time function $t$ on $ M$, the sets $S_a:= t^{-1} (a)$ and $J_a:= (I^-(p) \setminus J^-(q^{+}) ) \cap S_a $. First of all, the sets $J_a$ are nonempty, for all $a <t(p)$: If $I^-(p)$ and $J^-(q) $ have empty intersection in $I^+(S_a)$ then there is nothing to show. If they do intersect, then we choose $y \in I^-(p) \cap \partial J^-(q^{+}) \cap I^+(S_a)$. By compactness of $J^-(q^{+})\cap J^-(S_a)$ we can use elementary neighborhoods to show that there is $x \in S_a \cap \partial J^-(y) \cap \partial J^- (q^{+})$. Thus $ x \in J^- (I^-(p)) = I^-(p)$ by the push-up lemma. Openness of $I^+(p) $ and the fact that $x \in \partial J^- (q^{+})$ show that $ J_a$ is nonempty, containing an element $z_a$. Now choose past causal curves $ c_n$ from $p$ to $z_n$. The limit curve lemma implies that there is a limit curve in the closure of $M \setminus J^-(q^{+}) $, which is $M \setminus I^-(q^{+})$. As in turn $J^-(q)$ is a spatially compact subset of $I^-(q^{+})$ by Theorem 3 in \cite{oM}, the claim follows.      \hfill \qed

\bigskip

\V This concludes the proof of the theorem as well. \hfill \qed

\bigskip

\V Our requirements for conformal extensions are still strong enough to recover the usual prominent role of null geodesics at the boundary:

\begin{Lemma}
\label{ConvexFutureBoundary}
Let $E:(M,g) \rightarrow (N,h)$ be a conformal extension. Then $\partial^+ E(M) \cap \partial^- E(M)  = \emptyset$, and for all $ p,q \in \partial^+ E(M) $ with $ p \in J^-(q) $, there is a null geodesic curve $c: [0;1] \rightarrow \partial^+E(M) $ from $p$ to $q $ with $J^+(p) \cap J^-(q) = c([0;1])$. 
\end{Lemma}

\V {\bf Proof.} Choose $p_- \in E(M) \cap I^-(p)$. As $\ov{E(M)}$ is future compact, $J^+(p^-) \cap \ov{E(M)} $ is compact and contains $\ov{M_-}$ for $M_- := I^+(p_-) \cap E(M)$, thus the latter is open and precompact. Now assume $k: p \leadsto q$ is future causal intersecting $E(M)$. Then there is a future timelike curve $k_+ :[0;1] \rightarrow N$ with $k_+ (0) =p, k_+(1) \in \partial^+ E(M)$ and still intersecting $E(M)$. As shown in \cite{oM15c}, if $(M,g)$ is g.h. and $A \subset M$ is open and precompact, then $A$ is causally convex if and only if $\partial^+ A $ is achronal. Now apply this to $A= M_-$. As we saw that $\partial^+ M_-$ is not achronal, $M_-$ cannot be causally convex, thus $\ov{E(M)}$ is not causally convex, contradiction. Thus $c([0;1]) \subset \partial E(M)$. The rest of the argument is standard. \hfill \qed  

\bigskip

\V As it is to be expected, also in our context finiteness of the conformal factor $\Omega$ can be detected via null geodesics: 

\begin{Theorem}
\label{TheEndPointRules}
Let $(M,g)$ have a strong conformal extension $E: M \rightarrow N$ of order $\a \leq 2$. Let $c: \R \rightarrow M$ be a $C^0$-inextendible null geodesic in $(M,g)$. Then $c$ is of finite affine length if and only if for $q_c: =  \lim_{t \rightarrow \infty} ( E \o c ) (t) \in N $ we have $\omega^\a (q_c) \neq 0$.
\end{Theorem}

\V {\em Proof.}  One direction is easy: Obviously, if a $C^0$-inextendible null geodesic in $(M,g)$ has {\em infinite} affine length then $\omega^\a (q_c) =0$. For the other direction, define $k:= E \circ c$. It is a classical fact that $k$ is a lightlike pregeodesic in $N$, and the affine parameters $s$ of $c$ and $\hat{s}$ of $k$ are related by $ds = \omega^{-2} d\hat{s}$. Now assume that $q_c$ is an infinity point, i.e. $ \omega(q_c) =0 $.  For simplicity parametrize $k$ in the past sense, with $k(0) = q_c$. Then, as $\omega^\a \in C^1(N )$, we have that $u^\a:= \omega^\a \circ k$ is $C^1$ as well, and on any finite interval $[0,T] \ni t$ we have $u^\a (t) \leq A t  $ for an appropriate $A>0$, or, equivalently, for $B:= A^{1/\a}$, we have $u \leq B t^{1/\a}$. Thus the affine future length $l$ of $c$ can be calculated as $l = \int_0^T u^{-2} (s) ds \geq \int_0^T B^{-2} s^{-2/\a}  ds \geq B^{-2} \int_0^T s^{-1} ds=  \infty$, which implies the claim.  \hfill \qed

\bigskip

\V Theorem  \ref{ExtendibilityIsAsymptotic}, which we prove now, shows the abundance of conformally extendible spacetimes in Einstein-Maxwell Theory. Its Lagrangian density is $L: C^\infty (\tau^* M \x_{{\rm sym}} \tau^*M ) \times \Omega^1 M \rightarrow \Omega^n M$ is $ L(g,A) := {\rm scal}^g \cdot {\rm vol}^g + F \wedge * F$ for $F:= dA$. The corresponding Euler-Lagrange equations are

\begin{eqnarray*}
 d^*dA =0,\ {\rm Ric}^g - \frac{1}{2} {\rm scal} \cdot g = T(g,A)
\end{eqnarray*}

\V with $ T (g,A) (e_i,e_j) = \frac{1}{4 \pi} \big( \sum_{a} (F_{ai} F^a_j) - \frac{1}{4} \sum_{ab} (F^{ab} F_{ab}) g_{ij} \big)  $.

\V In this context, it is interesting to recall that Maxwell theory and more generally Dirac-Higgs-Yang-Mills (DHYM) theories have a nice conformal behaviour: If $ (\psi, \phi, A)$ is a solution of DHYM theory on a g.h. manifold $(M,g)$ then $(\Omega^{3/2} \psi , \Omega \phi, A)$  is a solution of DHYM theory on $(M,\Omega^{-2}g)$ (modulo the usual identification of spinors for two conformal metrics). If we restrict to {\em constant} conformal factors $\Omega (x) = c \in \R$, i.e., pure rescaling, and to Maxwell theory, then the Einstein tensor is invariant under scaling of the metric, whereas the energy momentum tensor $T$ of Maxwell theory has conformal weight $-1$ (inverse of that of the metric), and is quadratic in the curvature tensor $F=dA$, thus $T(c^2g, cA) = T(g,A)$, so for every solution $(M,g,A)$ of Einstein-Maxwell theory, $(M,c^2g, cA)$ is another solution. 

\bigskip

\V Einstein-Maxwell initial values $(S,g_0, K_0, A_0, \dot{A}_0)$ are called {\em Zipser-asymptotically flat (of mass $m$)} iff ${\rm tr}^{g_0}(K_0)=0$ and
\begin{itemize}
\item $(g_0)_{ij} \in (1+2m/r)\partial_{ij} + o_4 (r^{-3/2})$, 
\item $(k_0)_{ij} \in o_3(r^{-5/2})$,
\item $(F(A_0, \dot{A_0}))_{ij} \in o_3 (r^{-5/2})$.
\end{itemize}

\V for coordinates in a chart at infinity.

\V Zipser's theorem states that Zipser-asymptotically flat initial values satisfying a global smallness assumption have a causally complete maximal Cauchy development that can be foliated by maximal Cauchy hypersurfaces. The assumption of maximality (${\rm tr}^{g_0}(K_0)=0$) is merely a technical assumption that could be renounced by invoking the existence of a foliation by maximal hypersurfaces and an application of an inverse function theorem. 

\V The {\em global smallness assumption} for initial values is defined in \cite{nZ} as follows: For a complete initial datum $(g_0,K_0)$ we define for $a>0$ and $d_0$ being the $g_0$-geodesic distance to a fixed point:

\bean
 Q_a (g_0, K_0, F_0) := & a^{-1}& \big( \int_S ( \vert K_0 \vert^2 + (a^2 + d_0^2 ) \vert \nabla K_0 \vert^2 + (a^2 + d_0^2)^2 \vert \nabla^2 K_0 \vert ^2) d {\rm vol}_{g_0}  \\
 & +& \int_S ( \vert F_0 \vert^2 + (a^2 + d_0^2 ) \vert \nabla F_0 \vert^2 + (a^2 + d_0^2)^2 \vert \nabla^2 F_0 \vert ^2) d {\rm vol}_{g_0}  \\
 &+& \int_S (( a^2 + d_0^2 ) \vert \Ric \vert^2 + (a^2 + d_0^2)^2 \vert \nabla \Ric \vert^2 ) d {\rm vol}_{g_0} \big) 
\eean 
 
\V and $Q(g_0,K_0, F_0) := \inf \{ Q_a (g_0, K_0, F_0) \vert a>0 \}$. With these definitions, $Q(e_n, 0,0)=0$ where $e_n$ is the Euclidean metric, and $Q$ is continuous in $(e_n,0,0) + W^{3,2} \times W^{2,2} \times W^{2,2}$ where the Sobolev spaces are defined by means of the Euclidean metric.

\bigskip

\V {\em Proof of Theorem \ref{ExtendibilityIsAsymptotic}.} We want to apply \cite{hF} (for the higher-dimensional analogues of this result see Theorem 4.1 and Theorem 6.1 of \cite{mApC} and \cite{CBCL}) to the hyperboloidae of the maximal Cauchy development in coordinates given by a maximal temporal function. Note that we cannot apply 4.1 directly to the initial Cauchy hypersurface $S$, as the positive mass theorem forbids a $C^2$ extension (cf. Theorem \ref{X}). What we can do, however, is to consider the sequence of 'hyperboloidal' subsets $S_k := \{ x \in M \vert - (t (x)+k)^2 + r^2 = k^2 \} $ w.r.t. the coordinates $r,t$ defined by Zipser. Using the boundedness of the weighted quantities $W,K,L,O$ in Zipser's article, it is lengthy but straightforward to see that suitable conformal multiples of the metrics are indeed $C^l$ extendible, allowing for a conformal extension of $D^+ (S_k)$. For strongness it is important that we can use extension operators in the fashion of \cite{oM-BGB}, as we have to control not only the boundary but an open neighborhood of it. Uniqueness of the associated symmetric-hyperbolic system in the harmonic-coordinates constant-scalar-curvature (hccsc) gauge implies that all these conformal extensions coincide (in the chosen constant scalar curvature and harmonic gauge) on the intersection of their domains of definition.  To show the required behaviour at $i_0$, we use that there is an $\e >0$ s.t. if initial value $a$ has $\Vert a \Vert_{C^k_w} < \e  $ for a weight $w$ then the maximal solution has the required behaviour. Choose a higher weight $u$ and an associated weighted space $C_u^k$, then every initial value $b$ with $\Vert b \Vert_u^k < \infty$ satisfies that there is a compact subset $K \subset S$ such that for $b_{\rm ext}:=  b \vert_{S \setminus K} $ we have $\Vert b_{\rm ext} \Vert_{C_w^k} < \infty$. Complement $b$ on $ K$ to an initial value $B$ s.t. $ \Vert B \Vert_{C^k_u} < \e  $. Then the maximal Cauchy development of $B$ has the required end behaviour, and by local uniqueness of Einstein-Maxwell equations, its end is isometric to  $ D(S \setminus K  )  $ as a subset of the maximal Cauchy development of $b$. 

\V Causal convexity of $\ov{E(M)} = \ov{D(E(S))} $ is automatical, for strongness consider the symmetric hyperbolic equation satisfied by $\omega= \Omega^{-1}$ again in hccsc gauge and use the conventional stability arguments. Note that in the Penrose compactification of Minkowski space in the Einstein cylinder $(\mathbb{E}= \R \times \mathbb{S}^3 , - dT^2 + g_{{\rm round}})$ with the image $\{ x \in \mathbb{E} \vert 0 \leq \a (x) < \pi, \a (x) - \pi < T (x) < \pi - \a (x) \} $, we have $\omega = \beta^2$ for $\beta:=  \cos \a + \cos T = 2 \cos \frac{\a +T}{2} \cos \frac{\a - T}{2}$, thus $C^2$-stability in the gauge given by harmonic coordinates and constant scalar curvature of $N$ as chosen in the Penrose embedding in standard coordinates implies the claim. \hfill \qed

\bigskip
\V If $(M,g)$ admits a strong conformal extension $E: (M,g) \rightarrow (N,h)$ strong at a point $p \in \partial^+ E(M) $, then $(M,g)$ is isometrically extendible at $p$, i.e. there is $(Q,G)$ g.h., an isometric embedding $P: (M,g) \rightarrow (Q,G)$ and a point $q \in Q$ with $P^{-1} (I^-(q)) = E^{-1} (I^-(p))$. Of course, $(Q,G)$ is in general not the metric part of an Einstein-Maxwell solution. Indeed, it is a natural and interesting question whether for such Einstein-Maxwell solutions $(M,g,A) $ whose metric part admits a conformal extension, the electromagnetic potential $A$ can also be extended to $Q$. Such a statement would have a chance to be true only if one at least admits the application of a gauge transformation to $A$, asking: If $(M,g)$ admits a conformal extension $E$ yielding a conformal factor $e^{2f}$, does one find a regular extension of a gauge equivalent to $A$ to an open neighborhood of the closure of $E(M)$? The question is nontrivial because whereas $4$-dimensional Maxwell theory is conformally invariant, Einstein-Maxwell theory is not: A short look at the conformal transformation laws for $\Ein$ and $T$ reveals e.g. that at a point $p$ where $F(p)=0$ and $f(p)=0$, $d_p f = 0$ as well as $\nabla df \neq \l g$ for all $\l \in \R$, we know that $(e^{2f} g, A)$ is not an Einstein-Maxwell solution around $p$. One should expect Lorentzian analogs of the Riemannian removable-singularity theorems by Uhlenbeck (see \cite{kU}, \cite{tP}) to be helpful, but those would have to be proved by quite different methods as one cannot use elliptic regularity arguments any more. The canonical way would be to look at the hyperboloidal data including $A$ induced by the initial data and then invoke a theorem like the main theorem of \cite{mApC} but shifted from vacuum Einstein solutions to electrovac solutions, which is still to be elaborated (the claim does not follow from the application of \cite{mApC} together with a simple Kaluza-Klein correspondence, due to the lack of asymptotic flatness in the fifth coordinate and also as the result in \cite{mApC} is confined to even dimensions). Still, it seems plausible that {\em for any Einstein-Maxwell solution $(M,g,A)$ with initial values Zipser-close to trivial ones whose metric part is conformally extendible there is a gauge transformation $A \mapsto A + da= \tilde{A}$ such that $(M,g, \tilde{A})$ admits a conformal extension.} However, as we do not need the answer to the question above within this article, we are not going to pursue it further here.

\bigskip

\V The most standard vacuum solutions in relativity is certainly {\em Kruskal spacetime} $K_m$, being the manifold $\{ (T,X) \in \R^2 \vert - T^2 + X^2 > - 1   \} \times \SSS^2 $ equipped with metric $g$ given by $ \frac{32 m^3}{r} e^{-r/2M} \cdot (-dT^2 + dX^2) + r^2 (d  \theta^2 + \sin^2 \theta d \phi^2)$ where $r$ is the unique positive number such that $- T^2  + X^2 = (1 - \frac{r}{2M}) e^{r/2M}$, which is solved by $r= 2m(1 + W(\frac{X^2- T^2}{e}) )$, where $W$ is Lambert's W function. The subset $L_m$ of $K_m$ where $T> -X$ is called {\em Lemaitre spacetime}, whereas Schwarzschild spacetime $S_m$ is the subset of $L_m$ where $T <X$. It is easy to see that the Kruskal metric is (even locally) {\em isometrically} inextendible, by observing that along every inextendible null geodesic of finite affine parameter, the Kretschmann scalar $\langle R, R \rangle$ tends to infinity. But what about {\em conformal} extensions? It is well-known that $S_m$ does have a 'partial conformal extension' to the the exterior, with a smooth null boundary. A famous obstruction for conformal extensions to null boundaries are the peeling estimates due to Sachs \cite{S} refined by Newman and Penrose \cite{NP}, using null congruences. Other obstructions have been found in \cite{pc06} and \cite{oM15c}, as well as sufficient conditions in \cite{cL}, none of which is applicable to the question of the existence of a a {\em spacelike} future conformal boundary of the black hole $L_m \setminus S_m$ of $L_m$. Nevertheless, in Section 3 we will conclude that even locally, there is no such conformal extension.

\V Finally, it appears worthwhile to compare the above results to those obtained in \cite{cL}, \cite{LV}.

\section{Visual compactness and a variant of Hawking's singularity result}
\label{SCOTs}

\V In this section, we want to use the dominant-energy condition to connect the conformal structure of the spacetime to blackness of subsets. Let, for a pseudo-Riemannian submanifold $S$ of a pseudo-Riemannian manifold $M$, and for $X,Y \in T_pS$ and $N$ being the normal projection, let $\SecondFF (X,Y) := N(\nabla^M_X Y) = \nabla^M _X Y - \nabla^S_X Y$ and $H:= \tr (\SecondFF) = \sum_{i=1}^s \e_i \SecondFF(e_i, e_i)$ for a pseudoorthogonal basis $e_i$ of $T_pS$. We need a deformation lemma for maximal hypersurfaces appearing already in a proof of Burnett \cite{gB}:

\begin{Lemma}
\label{Deforming}
Let $\e >0$, let $S$ be a maximal partial Cauchy hypersurface of a globally hyperbolic TCC manifold $(M,g)$ and let $C \subset M$ with $S \cap C$ precompact. Then there is a hypersurface $S'$ with the properties:

\begin{itemize}
\item{$S' \subset J^+(S)$ and $d(S, S')< \e$,}
\item{$\partial S' = \partial S$,}
\item{$g(H_{S'} \vert_{C \cap S'},n) < 0$ where $n$ is the future normal vector of $S'$.}
\end{itemize}
\end{Lemma}

\V {\em Proof.} We consider a normal evolution $F$ of the inclusion $j$ of $S$, that is, a map $F: [0,1] \times S \rightarrow M$ with $F \vert_{\{ 0\} \times S} = j$ and such that $\partial_t F (t, s)$ is pointwise a multiple of the normal vector field $  \nu_{f_t(S)}$, i.e., there is $h \in C^\infty([0,1] \times S)$ s.t. $\nu = h \cdot F_{*} \partial_t$. Then

$$\frac{\partial H}{\partial t } = - h \big( \langle \SecondFF , \SecondFF \rangle + \ric (n,n) \big) + \Delta h .$$  

\V The timelike convergence condition implies that for $h>0$ the right-hand side is nonnegative for all $h$ with $\Delta h \leq 0$. Now we choose an open precompact subset $K \subset S$ with smooth boundary such that $C \cap S \subset K$. Choose $u(s) = h (t=0, s)$ such that $\Delta u = - f \in C^\infty (S) $ with $f >0$ on $K$ and $u=0$ on $\partial K$. Hopf's maximum principle implies that $u>0$ on $K$, thus $f \geq c>0$ on $C \cap S$. Thus $\frac{\partial H}{\partial t} \leq - c$ on $C \cap S$, and, consequently, for small $t$, we have $H(S_t \cap C) <0$. \hfill \qed    

\bigskip
\V Let $(M,g) $ be a spacetime and let $P(M)$ be the power set of $M$. We define two maps $U_\pm: P(M) \rightarrow P(M) $ by $ U_- (A) := J^+(J^-(A))$ and $U_+ (A) := J^-(J^+(A))$ for every $A \in P(M)$. Obviously, the maps $ U_\pm$ are monotonous w.r.t. the order given by inclusion, and $A \subset U_\pm (A)$ for all $A \subset M$. A subset $A$ of $M$ is called {\em $n$ times future (resp. past) visually compact} iff $U_+^n (A)$ (resp. $U_-^n (A)$) is spatially precompact, obviously for any $n \in \N$, any $(n+1) $ times visually compact subset is $n$ times visually compact, and a $1$ time visually compact subset is simply called visually compact. Note that if $(M,g)$ is conformally extendable and a closed subset $A$ of $M$ is visually compact then there is an open neighborhood $U$ of $A$ that is still visually compact. We call $p \in M$  future resp. past visually compact iff $\{p\}$ is. Note that for a precompact open set $U$, if $J^-(J^+(\partial U))$ is spatially precompact, then also $J^-(J^+ (\ov{U}))$ is spatially precompact. Note also that visual compactness is a conformally invariant notion (while blackness is not). However, Theorem \ref{VariantHawking} shows that the TCC connects the two notions.

\bigskip

\V {\em Proof of Theorem \ref{HorizontalCITIFs}.}  
\V Let $S$ be such a maximal hypersurface in $I^-(c)$. There is a Cauchy hypersurface $\Sigma$ that contains $S$ as an open subset. Theorem \ref{Deforming} implies that $S$ can be deformed to a hypersurface $S'$ with $\partial S' = \partial S$, $d (p, S') < 1 $ and $\partial S' \cap J^- (U(p)) $ has past mean curvature vector. As $J^- (c)$ is spatially compact, $H\leq -b <0$ on $J^-(c(I)) \cap S'$. Obviously, $I^-(c) \cap I^+ (\Sigma) \subset D^+(S) = D^+(S')$.  But then Hawking's singularity theorem applied to the globally hyperbolic manifold $D(S')$ with trapped Cauchy surface $S'$ implies that any curve from $S'  $ has length $< b^{-1}$, so $l(c) < b^{-1} + 1$.  \hfill \qed

\bigskip

\V As mentioned in the introduction, the assumption of the existence of a maximal hypersurface (actually a foliation by such, in the entire spacetime) is satisfied for the maximal Cauchy development of an arbitrary element of a neighborhood of trivial initial data, and the intersection of any of those with $I^-(c)$ yields a maximal hypersurface with the boundary condition in the previous theorem. For initial data far from the trivial ones, we describe in the following a method to construct such a maximal hypersurface if there is a strong extension.


\bigskip

\V {\em Proof of Theorem \ref{VariantHawking}}. Let $ x \in L:= J^-( J^+(p)) \cap \partial^+ E(M)$ be given. As $J^-(J^+(p))$ is spatially compact, future-compactness of $\ov{E(M)} $ implies that $ J^-(x) \cap \partial E(M) $ is compact, and thus its boundary in $\partial^+ E(M) $ consists of non-dominating points. Then we proceed as in the proof of Theorem \ref{HorizontalCITIFs}. \hfill \qed

\bigskip

\V We now want to prove Theorem \ref{EmergingMaximal}, but before let us remark that close to trivial initial data we have $s=1/2$, {\em not} satisfying the condition in the theorem. On the other hand, the growth condition on $\Omega$ needed in the proof could, instead of assuming $s>\frac{n-1}{4}$, also be derived if there is a conformal $C^k$ extension such that $\omega^{\frac{n-1}{4}}$ has a $C^k$ extension with $\grad  \omega (p) = 0$ and an additional open condition on the Hessian of $\omega$ at $p$ is satisfied. And indeed, in the Penrose compactification, for $\beta := \sqrt{\omega}$, the Hessian of $\beta$ is a constant multiple of the metric on the orthogonal complement of the spherical orbits and thus ${\rm Hess} (\omega) = 2 \beta {\rm Hess} (\beta) + d \beta \x d \beta $ is positive definite in all null directions not tangential to the boundary in a neighborhood around $ J^+(i_0) = (T+\a) (\pm \pi)$. Another example of a spacetime satisfying the growth condition is Schwarzschild spacetime, by the appropriate dimensional reduction mentioned in the previous section.

\bigskip

\V {\em Proof of Theorem \ref{EmergingMaximal}.} We want to construct a maximal hypersurface in $I^-(c)$ by constructing barriers. We parametrize $c$ by arclength and with $c(0) = p$. Let $r_n$ be an $S $- maximal geodesic curve from $p$ to $c(n)$. The limit curve lemma and the fact that $c$ has a spatial endpoint imply that the sequence $n \mapsto r_+^n$ of causal curves has a limit curve $r_+$ that is a maximal future ray and $I^-(r_+) = I^-(c)$: Assume that there is $s \in \R$ with $r_+(\R) \cap I^+(c(s)) = \emptyset$. As $A:= J^-(c)  \setminus I^+(c(s))$ is compact, we can choose a complete Riemannian metric $G$ on $I^-(c) $ coinciding with a flip metric on $A$ w.r.t. any Cauchy temporal funktion $\tau$, and as the metric coefficient in the $\tau$-splitting in front of $d \tau^2 $ is bounded, there is a universal bound $D$ on the $G$-length of all causal curves in $A$. The Limit Curve Lemma then implies that eventually members of the sequence of curves enter $I^+(c(s)) = I^-(c) \setminus A$ after the parameter $D+1$. This shows $I^-(r_+) = I^-(c)$. Moreover, $r_+$ is timelike, as can be seen by $S$-maximality of the segments and the choice of an appropriate open neighborhood of $c([0, \infty))$.
 
\V Recall that the future ray horosphere $S^-_\infty$ of $r_+$ in the terminology of \cite{GV} is defined by $S^-_\infty := \partial \big( \bigcap_{k \in \N} J^- (S_k^- (r_+(k))) \big)$, where $S_k^-$ is the Lorentzian past sphere of radius $k$. The idea is to use $S^-_\infty$ as a barrier surface to obtain a maximal hypersurface. Following a slight modification of Theorem 4.2 in \cite{GV}, $B^-:= S^-_\infty$ has mean curvature $\geq 0$ (w.r.t. the future normal) in the support sense, that is, for all $q \in B^-$ and all $\e>0$, there is a $C^2$ spacelike hypersurface $S_\e$ such that there is a neighborhood $U$ of $q$ in which $B^-$ is acausal and edgeless, $S_\e \cap U \subset J^-(B^-,U)$. The slight modification mentioned above (apart from time-dualizing) consists in the fact that to obtain the conclusion of the theorem, one does not need completeness of the spacetime but only future completeness of the normal geodesics of $S_\infty^-$ --- which is satisfied in our case, as every point is the limit of points of past spheres of points on $r_+$ --- and the same proof works. Note that $B^-$ satisfies all past barrier properties: Lemma 3.22 of \cite{GV} ensures that $B^- \subset J^-(S) $, Theorem 4.2 of \cite{GV} states that its mean curvature is $\geq 0$ in the support sense, $D^-(S) \cap J^+(B^-)$ is compact (because $J^+(J^-(x)) \cap J^-(S) $ is compact and consists of non-black points), $S \subset I^+(B^-)$, and $ \partial (D(S) \cap I^+(S^-) )  = (\partial (D(S)) \cap I^+(S^-)) \cup (S^- \cap D(S)) $.

 Using the nowadays classical results on convergence of CMC flows in spacetimes in the case of barriers (see e.g. Theorem 2.1 in \cite{kE} or the preceding articles by Ecker-Huisken or Gerhardt), translated to the Dirichlet case in Volker Christ's MSc thesis \cite{vC}) (in particular Theorem 4.1 and Theorem 4.2 of the thesis), we can find a maximal surface $F$ in (the region of $\hat{M} $ isometric to) $M$ with $D(F) \supset U_+(x)   $ via the following a priori estimates. The a priori height estimate towards the past is given by a barrier, and a small modification in comparison to the usual arguments is that we admit a subset with $H \leq 0$ {\em in the support sense} as a barrier, but the argument that the flow cannot touch that subset works verbally the same. To obtain a height estimate towards the future, instead of a barrier we use another argument for why the flow $\Phi_s$ does not converge to the future lightlike boundary of $I^-(c) = I_N^- (i_+)$. This is done in the following by giving a-priori estimates for $t \o \Phi_s$, where $t$ is a temporal function on the conformal extension $(N,h) \supset E(I^-(c))$ with $t(i_+) = 0$. We define $S_n:= E(I^-(c)) \cap t^{-1} (-1/n)$. Strongness of the conformal extension implies existence of a number $C>0 $ with $ \Omega < C t^\b  $ with $\b < 4/(n-1)$ for the conformal factor $\Omega$ on $E(I^-(c))$ (and this is indeed the only fact we need here from the conformal extension $E$). On one hand, $\Phi_s$ is a smooth hypersurface for all $s>0$ with $s \mapsto \vol^{(n-1)} (\Phi_s)  $ monotonously increasing. On the other hand, let, for $n \in \N$ and $f \in C^1(E(I^-(c)), (0; \infty) )$,
 
 \bean
 V_\Omega(n) := \{ \sup \vol^{(n-1)}_{\Omega \cdot h} A  \vert &A {\rm \ spacelike \ } C^2-{\rm hypersurface \ in \  } E(I^-(c)) {\rm \ with \ } \partial A = \partial \Sigma \\
 &{\rm and \ } A \cap t^{-1} ((-1/n;0)) \neq \emptyset \}. 
    \eean
 
\V  Now we zoom into an appropriate neighborhood $U$ of the tip covered by $h$-Riemann normal coordinates at the tip. In the following estimates, we could then replace the Minkowski content by the usual volume. We however prefer the former, to make the lemma applicable also for large, possible noncompact, subsets of the future boundary, where no uniform estimate of the injectivity radius is available.  
 
 \begin{Lemma}
Let $S_k \subset U$ where $U$ is as above. There is a number $C>0$ such that for all $N \in \N$ with $N>k$ and all $q_N \in S_N $ we have 
 
 $$ v_{n,k}^\Omega:= \vol^{(n-1)}_{\Omega h} \big( E(I^-(c) \setminus J^-(q_N  )) \cap S_k \big) < C \cdot  \Omega^{\frac{n-1}{2}} \cdot N^{-1} \cdot k^{-2}   . $$
 \end{Lemma} 
 
 \V {\em Proof of the lemma.} Let $Q_k:= \partial J^-(i_+) \cap S_k$, then Clarke's inverse function theorem for Lipschitz maps implies that $Q_k$ is a Lipschitz $(n-2)$-manifold, and as such its $(n-2)$-Minkowski content $M_{n-2} (A)$ exists, where 
 
 $$  M_{n-j} (A) =  \lim_{r \rightarrow 0^+} \  (\mu (B_r (A)) / \a (j) r^j)  $$
 
 \V  and where $\a(j) r^j$ is the volume of the Euclidean ball of radius $r$, in particular $\a(2) := 1$, see standard textbooks on geometric measure theory, e.g. \cite{Federer} or \cite{KrantzParks}. On the other hand, Gr\"onwall's Lemma yields existence of real numbers $C'$, $C''$ with 
 
 $$  \vol^{(n-2)} (Q_k) < C' \cdot \frac{1}{k}, \qquad \d_\Omega (N) := d_{(S_k, \Omega h)} (\partial J^- (q_N) , \partial J^-(i_+)) < C'' \cdot \frac{1}{N} ,$$
 
\V which together with the previous statement implies the claim of the lemma. \hfill \qed

 \bigskip
 
 \V Now we calculate 
 
 \bea
\label{Dreiecksflaechenabschaetzung}
 V_\Omega (n) < \sum_{k=1}^N v_{k,N}^\Omega < \frac{C}{N} \sum_{k=1}^N  (1/k)^{-\beta} \frac{1}{k^2} < \frac{C}{N} \sum_{k=1}^N k^\g \rightarrow_{N \rightarrow \infty} 0 , 
 \eea

\V with $\g <0$, which means that, by monotonicity of the volume under the flow, some $S_N$ will never be intersected by the flow, implying the last necessary a-priori estimate preventing that the flow converges to the lightlike future boundary of $I^-(c)$, ensuring that it converges to a maximal hypersurface with the given boundary. \hfill \qed


\bigskip

\V Which spacetimes contain spatially precompact TIPs? Schwarzschild and Kruskal spacetime do --- their black points are exactly those from which only future curves $c: I \rightarrow M$ emanate with $I^-(c(I)) $ spatially precompact, as can be seen most directly by diagrams using a notion developed by Chru\'sciel, \"Olz and Szybka \cite{COS} making rigorous the previously rather vague idea of 'Penrose diagram': Let $(M,g)$ be a spacetime, then a {\em projection diagram}\footnote{We specialize this notion to $U=M$ in the terminology of \cite{COS} as we are primarily interested in globally hyperbolic manifolds and in black hole spacetimes which typically have the problematic symmetry axis $r=0$ at the boundary of the diagram.} of $(M,g)$ is a submersive map $\pi \in C^\infty(M, \R^{1,1})$ mapping timelike curves to timelike curves (so the tangent space of the fiber does not contain timelike vectors) and such that for every timelike curve $c$ in $\pi(M) \subset \R^{1,1}$ there is a timelike curve $C$ in $M$ with $c= \pi \circ C$. Every conformal submersion $F: M \rightarrow \R^ {1,1}$ with complete Riemannian fiber satisfies this assumption. The article \cite{COS} then gives examples of projection diagrams: First, the usual triangle diagram is a proper projection diagram of $\R^{1,3} \setminus \{r=0 \}$, whereas the usual two-dimensional diamond is a projection diagram for the entire spacetime $\R^{1,3}$ but then the projection is not proper, as we mod out the translational symmetry of $\R^2$ instead of the rotational one, where the fiber was $\S^2$. Here we are less interested in properness of the projection (compactness of the fibers) than in the {\em completeness of the fibers}: If all fibers are complete, we call the projection diagram {\em complete} (of course, properness implies completeness). This is the case for the diamond projection above but not for the triangle projection. If $\pi$ is a complete diagram, then global hyperbolicity of $\pi(M)$  implies global hyperbolicity of $M$. We call a projection diagram {\em embedded} if there is a conformal embedding of $\pi(M)$ with the metric induced from $\R^{1,1}$ into $(M,g)$ which is a right inverse of $\pi$. Let us call a spacetime SCOT if it contains spatially precompact TIPs. Then if $(M,g)$ is SCOT and $\pi$ is a projection diagram for $(M,g)$, then $\pi (M) $ is also SCOT. If $\pi$ is complete in the sense above, then SCOTness of $(M,g)$ is equivalent to SCOTness of $\pi (M)$. In \cite{COS} it is shown that, for Schwarzschild spacetimes $m>0, a=0$, the usual half-hexagonal Penrose diagram is indeed a proper embedded projection diagram, and for $a>0$ it is shown: 

\begin{Theorem}[see \cite{COS}]
\label{COS}
For every element of the entire slow Kerr-Newman family \footnote{Here we refer to the case $a\leq m$ where $m$ is mass and $a$ is angular momentum and to the maximal Cauchy developments of the initial values with one incomplete inner end and one outer asymptotically flat end. If we chose a slice with  two asymptotically flat ends instead, every maximal Cauchy developments would consist of four blocks instead of two.} with $a>0$ there is a proper embedded projection diagram onto a causal diamond in $\R^{1,1}$. One of the two null geodesic segments whose union is the future boundary of the diagram represents the inner, Cauchy horizon consisting of endpoints of curves of finite length. The part $n_1$ between $i_0$ and $i^+$ of the other null geodesic segment $n$ is future null infinity, whereas the part $n_2$ of $n$ to the future of $n_1$ consists of ideal endpoints of curves of finite length.  \hfill \qed
\end{Theorem}

\V Looking at the diagrams we get as a trivial corollary:

\begin{Theorem}
In the Kerr-Newman family, the members of the Schwarzschild family with $m>0$ are the only ones that contain spatially precompact TIPs. \hfill \qed
\end{Theorem}

\V A special case of a projection diagram is an {\em orbit diagram}. The latter arises by quotienting out a conformal group action whose orbits are $2$-codimensional. If the fibers are complete, then the two defining properties for projection diagrams are easily checked, even when dropping the codimensionality assumption, getting a submersive map $\pi: M \rightarrow N$ between spacetimes of arbitrary dimension. Also in this general case completeness of the fibers and global hyperbolicity of the image imply global hyperbolicity of $M$. The fibers are complete in particular if they are discrete (in the case of a free and properly discontinuous group action) or compact (as in the case of a spherical symmetry group).

\bigskip

\V As conformal maps map IPs to IPs, it is easy to see that in the above situation, if we denote the future conformal boundary of $M$ resp. $N$ by $\hat{M}$ resp. $\hat{N}$ (each equipped with the canonical metric given by the volume of the symmetric difference of sets w.r.t. a finite volume form, see \cite{oM15c}) and the canonical inclusions $M \rightarrow \hat{M}$ resp. $N \rightarrow \hat{N}$ by $i_M$ resp. $i_N$ then we can find a continuous surjective map $\hat{\pi}$ with $\hat{\pi} \o i_N =  i_M \o \pi$ (compare with \cite{lAlH}). However, even in the case of an isometric action of a group $G$, $\hat{\pi}$ is i.g. not a $G$-principal bundle, as one can see in the example mentioned in \cite{lAlH}: Let $M= \R^{1,1}$ and $G= \Z v$ for the action of a translation by a spacelike vector, then $N$ is a cylinder, and the fiber of the IP $N$ contains as well $\R^{1,1}$ as $(x_0 - x_1)^{-1} ((- \infty; 0))$, which is the past of a null geodesic. But those two sets are not related by a group element, actually by no conformal transformation of $\R^{1,1}$. The best one can say is that the group action is almost transitive on each fiber $F$ in the sense that $F$ is the closure of every orbit of $f \in F$.

\bigskip

\V Recall that the restriction of every conformal extension of a spacetime $(M,g)$ to the timelike future $I^+(p)$ of any $p \in M$ would be an open conformal embedding with a precompact image. The example of Schwarzschild spacetime shows that i.g. even spherical orbit diagrams for an isometric action do not correspond to conformally extendible spacetimes:

\begin{Theorem}
Let $p \in L_m$, then there is no open conformal  embedding of $I_{L_m}^+(p)$ into a spacetime  with precompact image.
\end{Theorem}

\V {\em Proof. } We show the statement for $p \in B:= L_m \setminus S_m$ which is no restriction of generality, as for every $q \in L_m$, we have $I^+(q) \cap B \neq \emptyset$. Now denote the spherical orbit projection to the two first components by $\pi$. As the spherical metric is in $O(r^2)$ whereas the $dX$ component is in $O(r^{-1})$, it is obvious that for every $C^0$-inextendible timelike future curve (CITIF for short) $c: [0, a) \rightarrow B $ we have $\lim_{t \rightarrow a} r(c(t)) = 0$ and therefore $ \exists b \in (0; a)$ with $\pi^{-1} (c(b)) \subset I^+(c(0))$. Consequently, we have 

\bea
\label{PastFactorizesThroughProjection}
I^-(c) = \pi^{-1} (I^-(\pi \o c)).
\eea

The future causal boundary $\partial^+ B$ is made up of all TIPs in $B$, which correspond to pasts of CITIFs by a classical result of Geroch,   topologized e.g. by the metric of volume of symmetric difference, see \cite{oM15c}. Because of Eq. \ref{PastFactorizesThroughProjection} and as the fibers of $\pi$ are compact, we get a homeomorphism $\Psi$ from the future causal boundary of $B$ to the one of $\pi(B)$. Now, as $\pi(B)$ is a two-dimensional g.h. spacetime and thus admits a conformal extension (see \cite{Spec}), its causal boundary is a one-dimensional topological manifold. Thus $B$ cannot have a conformal extension, as otherwise its future causal boundary would be homeomorphic to the future boundary of the conformal extension (see \cite{oM15c}), which would be three-dimensional. \hfill \qed

\bigskip

\V However, the following result can help to use projection diagrams as a source for counterexamples against naive conjectures:

\begin{Theorem}
\label{ConformalTCCinCompact}
\V Let $(M,g)$ be a cosmological globally hyperbolic spacetime with a Cauchy surface $S$ and a steep Cauchy temporal function $t$ such that ${\rm Hess} (t)$ is bounded on each level set of $t$, then there is $u \in C^\infty (M, \R)$ such that $\tilde{g} := e^{2u_\pm} g$ satisfies the TCC and $ u \vert_{J^+(S)} >0$.
\end{Theorem}

\V {\em Remark.} The required property of the Hessian is automatically satisfied by cosmological, i.e. globally hyperbolic and spatially compact, spacetimes.

\V {\em Proof.} We consider the classical formula 

$ \widetilde{\ric} = \ric - (n-2) (Ddu - du \x_{{\rm sym}} du ) + (\Delta u - (n-2) g(\grad^g u, \grad^g u) ) \cdot g $

for the Ricci curvature $\widetilde{\ric}$ of the metric $\tilde{g} := e^{2u} g$ (see for example Eq. 1.159 d) from \cite{BESSE}). Let $H_a$ be the compact set of causal vectors $v$ in $T t^{-1} (a)$ such that $dt (v) =1$, then every causal vector in $T t^{-1} (a)$ is a multiple of a vector in $ H_a$. Now we choose $u:= \psi \o t$ for a Cauchy temporal function $t$ and a strictly monotonously increasing function $\psi: \R \rightarrow \R$. Focussing on the quadratic terms $df \x_{{\rm sym}} df$ and $g(\grad^g u, \grad^g u) ) \cdot g$ we see that the latter is nonnegative on $H_a \times H_a$ whereas the former is even uniformly positive on $H_a \times H_a$, which implies that there is a sufficiently large constant $C_0 (a)$ such that the TCC is satisfied on $T t^{-1} (a)$ as soon as for $ \phi:= \psi'$ we have $\phi'< C_0 \phi^2 $. The function $a \mapsto C_0(a)$ is continuous, thus we can choose a smooth function $a \mapsto C (a) $ with $ C > C_0 $. Then for $a:= \ln  \phi$ we need $C \cdot a' < e^a$. Solving this inequality from a level set of $t$ towards $ \infty$ along the lines of \cite{MN} yields the desired functions $u$. \hfill \qed

\section{A variant of Penrose's singularity theorem}
\label{jacobi}

\V In this section we revise the usual prerequisites of the Penrose singularity theorem and derive a slightly different version. We first recall some classical facts about adapted Jacobi fields for the non-expert reader. Let $N_1, N_2$ be spacelike submanifolds of $M$ and let $ c : [0,E] \rightarrow M$ be a timelike geodesic from $p_1 \in N_1$ to $q \in N_2$ which is orthogonal to the submanifolds at the endpoints, i.e. $ \dot{c}(0) \perp_g T_p N_1  $ and $ \dot{c}(E) \perp_g T_q N_2  $. A Jacobi field $J$ along $c$ is called {\em $(N_1, N_2)$-adapted} iff $J(0) \in T_p N_1$, $ J(E) \in T_q N_2$, ${\rm proj}^{tan}_g(\nabla_t J(0) )= - \hat{S}^{N_1}_p (J(0) , \dot{c} (0))$ and ${\rm proj}^{tan}_g(\nabla_t J(E) )= - \hat{S}^{N_2}_q (J(E) , \dot{c} (E))$ where the $\hat{S}$ denote the second fundamental forms of the respective submanifolds seen as bilinear maps $TN_i \times T^\perp_g N_i \rightarrow TN_i $. More explicitly, for all $Y \in TN_1$ we have $\langle \nabla_t J(0) , Y \rangle = - \langle S^{N_1} (J(0) , Y) , \dot{c} (0) \rangle$. The adaptedness is closely related to a similar property of geodesic variations, namely {\em $(N_1, N_2)$-properness}. A geodesic variation $V$ of $c$ is called {\em $(N_1, N_2)$-proper} iff $V(s, 0) \in N_1$, $V(s,b) \in N_2$, $\partial_t V(s, 0) \perp TN_1$, $\partial_t V(s,b) \perp TN_2$ for all $s \in (-1,1)$. Analogously, in the obvious way, we introduce the notions 'initially adapted' and 'terminally adapted'. As an easy consequence of torsion-freeness of the Levi-Civita connection (pulled back to the domain of a variation) we get a well-known relation between the two notions:  The variational vector field of an $(N_1, N_2)$-proper geodesic variation is an $(N_1, N_2)$-adapted Jacobi field. Conversely, if a Jacobi field $J$ is $(N_1, N_2)$-adapted, then there is an $(N_1, N_2)$-proper geodesic variation with variational vector field $J$. If $t \in (0,E)$ is such that the dimension $A(N_1,  c, t)$ of the vector space of $(N_1, \{ c(t) \})$-adapted Jacobi fields is nonzero, $t$ is called {\em $N_1$-focal point to $0$ along $c$ (of order $A(N_1, c,t)$)}. It is well-known that after the first $N_1$-focal point, a timelike geodesic cannot be maximal between the endpoint and the initial submanifold.

\V We first revise a well-known intermediate theorem holding without causality assumptions on $(M,g)$. To formulate it, we first need to quantify affine length of a null geodesic by an initial gauge, which is defined via scalar product with some timelike vector field $V$ (which exists due to time orientation). Explicitly, for a spacelike codimension-two submanifold $A$, we set $B_A:= \{ v \in \tau M^{-1} (N) \vert v \perp TN , g(v,v) = 0, g(v,V) = -1  \} $. Note that this is a subbundle of $\tau M ^{-1} (N)$ with zero-dimensional fiber. The fiber consists of two points, and it is easy to see that if $N$ is a connected oriented submanifold then $B_A$ is disconnected. For example, if $A = \partial G$ for a spacelike hypersurface $G$ then we have an inner and an outer null normal, so $B_A$ is the disjoint union of $B_A^+$ (outer null normals, generated by $k_+ := \nu + n$, where $\nu$ is the future timelike normal of $G$ and $n$ is the outer normal of $A$ in $G$) and $B_A^-$ (inner null normals, generated by $k_- := \nu - n$). A spacelike codimension-two submanifold $A$ is called {\em outer trapped resp inner trapped} iff $ g(H_{A \subset M} , k_+)<0  $ resp. $g(H_{A \subset M} , k_-) <0$, and as $\nu=k_- + k_+$, obviously $A$ is trapped iff it is inner and outer trapped. A spacelike codimension-two submanifold $A$ is called {\em marginally outer resp. inner trapped} iff $ g(H_{A \subset M} , k_+)=0  $ resp. $g(H_{A \subset M} , k_-) =0$. An outer trapped surface resp. marginally outer trapped surface is abbreviated OTS resp. MOTS. Let us first, for greater self-containedness, revise the following complete analogue of a classical result:

\begin{Theorem}
\label{intermediate}
Let $(M,g)$ be an $n$-dimensional Lorentzian manifold and $A \subset M$ a spacelike $2$-codimensional submanifold of mean curvature $H$. Let $C:[0 , D) \rightarrow M$ be a null geodesic starting at $p \in A$ and $\dot{C} (0) \in B_A^+$. If there is $b>0$ with $ g(H(p) , \dot{C} (0)) \leq -1/b$ and $\ric(\dot{C} (t),\dot{C} (t)) \geq 0$ for all $t \in [0,b]$, then $C$ is not $A$-maximal after $ b$. Consequently, if $H$ is uniformly future, $J^+(A) \setminus I^+(A) \subset \exp (([0,b) \cdot B_A) \cap {\rm Def (exp)})  $.  
\end{Theorem}

\V {\em Proof.}  We first assume that $C$ is $A$-maximal beyond $ b$ and restrict $c:= C \vert_{[0, b]}$. It is an easy classical result (using the symmetry of the Riemann curvature tensor) that for any two nontangential Jacobi fields along a geodesic $k$, $g(\dot{J}_i , J_j) - g(J_i, \dot{J}_j)$ is constant along $k$. This implies the following well-known fact: Let $J_1, ... , J_l$ be notangential Jacobi fields along a geodesic $k: [0, E] \rightarrow M$. Then if there are functions $f_i$ on $[0,E]$ such that $X = \sum_{i=1}^l f_i J_i$, we have

$$g(\dot{X}, \dot{X}) - g(R(X, \dot{k}) \dot{k}, X) = g(L,L) + \frac{d}{dt}g(X, B)  $$

\V where $L := \sum_{i=1}^l \dot{f}_i J_i $ and $B:= \sum_{i=1}^l f_i \dot{J}_i$.

\V Now we apply this fact to $k=c$ and to $A$-adapted Jacobi fields $J_i$ along $c$ with $J_i(0)=e_i$ for an orthonormal basis $e_1, ... e_{n-2}$ of $T_pA$. The lemma implies that for every smooth vector field $X $ along $c$ with $X(E)=0$ and $X(t) \in span(J_1 (t),... , J_l (t))$ for all $t \in [0,E]$, we have 

$$\int_0^b ( g(\nabla_t X , \nabla_t X) - g(R(\dot{k}, X) X, \dot{k}) ) dt - g (X(0), \sum f_i \dot{J}_i ) > 0$$

\V where the $f_i$ are defined by $ X= \sum_{k=1}^l f_k J_k $, because $g(L,L) \geq 0$. Now we apply this to the non-tangential vector fields $X_i$ along $c$ defined as $X_i(t) := (1 - t/b) P^t_{c} (J_i(0)) $, obtaining $\nabla X_i (t) = - \frac{1}{b} P^t_{c} (J_i(0))  $, and $X_i(t) = \sum_{j=1}^{n-1}f_i^j (s) J_j (s) $, with $f_i^j \in C^\infty ([0,b])$ and $f_i^j (0) = \delta_i^j$, and $ g (X(0), \sum f_i \dot{J}_i ) = g (\dot{c} (0), S_N(X_i(0), X_i(0)) )$ due to $S$-properness. So we get 

$$0 < \int_0^b ( g(\nabla_t X_i , \nabla_t X_i) - g(R(\dot{c}, X_i) X_i, \dot{c}) ) dt - g (\dot{c} (0), S_N(X_i(0), X_i(0)) ) . $$

\V We choose $a=1$ and sum over $i$. Then we use the general fact that in an $n$-dimensional Lorentzian manifold $(M,g)$, for a null vector $v \in T_pM $ and $e_1, ... , e_{n-2} $ spacelike unit vectors that are orthogonal to each other and to $v$, we get 

$\sum_{i=1}^{n-2} g(R(v, e_i)e_i, v) = ric(v,v)$. And we calculate for $e_i:= P_c^t (J_i(0))$:

\bean
0 &<& \frac{n-2}{b^2}b  - \int_0^{b} (1 - \frac{t}{D_0})^2   \cdot \sum_{i=1}^{n-2} g(R(\dot{c}, e_i) e_i,  \dot{c} ) dt + (n-2) g(\dot{c} (0) , H(p) )\\
&=& \frac{n-2}{b^2} b + (n-2) g(\dot{c} (0) , H(p))  - \int_0^{b}  (1 - \frac{t}{b})^2   \cdot ric (\dot{c} (t) , \dot{c} (t) ) dt\\  
&<&  \frac{n-2}{b} + (n-2) g(\dot{c} (0), H(p) )
\eean

\V (where the last step uses the condition on the Ricci tensor), contradiction. \hfill \qed

\bigskip

\V It is worthwhile to compare this intermediate theorem to Proposition 1 of the beautiful recent paper \cite{GS}, which has been an approach to unify singularity theorems using trapped surfaces of different codimensions. The crucial condition of that proposition is $ \tr (g(R(\dot{c}, \cdot) \dot{c} , \cdot) \vert_{P \times P}) >0$, where $P$ is the parallel transported $TN$ along $c$. This condition does not follow from the timelike convergence condition, although the authors of \cite{GS} remark correctly (in Remark (ii) after that proposition) that in the case of codimension $2$ the condition does follow from the null convergence condition {\em if the curve is null} (an assumption made implicitly by the authors of \cite{GS} by their notion $n:= \dot{c} (0)$).

\begin{Theorem}
\label{TrappedInTheAct}
Let $(M,g)$ be a connected globally hyperbolic Lorentzian manifold and $A \subset M $ a compact achronal spacelike outer trapped codimension-two submanifold of $M$. If $\ric_M(X,X) $ for all null vectors in $J^+(A)$, then either $\partial^+ (D(\ext (A)))$ is compact, or there is a noncomplete null geodesic starting on $A$. Moreover, $\partial J^+(A) $ is ruled by incomplete null geodesics, and if $(M,g)$ admits a strong conformal extension $E$ of order $\a \leq 2$, then $\omega$ does not vanish on $Z:= \partial^+ E(M) \cap\partial  J^+(A)   $. 
\end{Theorem}

\V {\em Proof.} Compactness of $B_A^+$ implies that $g(H_M, \cdot)$ attains a positive minimum $1/b$ on $B_A^+$. As geodesics stop being $A$-maximal after the first $A$-focal point, either there is some incomplete null geodesic starting on $A$ (so that $\exp$ is not defined for a sufficiently long time) or each point $q \in \partial^+ (D(\ext (A)))$ is of the form $\exp_M (a \cdot v)$ for $v \in B_A^+$ and $a \leq b$, thus $\partial^+ (D(\ext (A)))$ is compact. The statements in the last sentence are due to the fact that the geodesics making up $\partial J^+(A)$ are either maximal on their entire domain of definition and thus of finite length or cease to be maximal and thus leave $\partial J^+(A) $ after a finite affine time. The very last statement follows directly from Theorem \ref{TheEndPointRules}. \hfill \qed

\bigskip

\V The next theorem is an immediate corollary and already appears similarly in \cite{AMMS}. However, later we will rather need the statement of the previous Theorem \ref{TrappedInTheAct}.  

\begin{Theorem}
\label{VariantPenrose}
If $(M,g)$ satisfies the null energy condition and contains a noncompact Cauchy surface, which contains a compact subset $U$ whose boundary is an outer trapped surface, then there is a $C^0$-inextendible incomplete null geodesic from $\partial U$.  
\end{Theorem}

\V {\em Proof.} First we use the result by Bernal and S\'anchez \cite{BS} that every compact spacelike acausal submanifold-with-boundary (as $U$ is one) is a subset in a smooth Cauchy hypersurface to find a Cauchy hypersurface $S$ of $M$ containing $U$. For $\ext (A) := S \setminus U$, $ F:= \partial^+ (D^+(\ext(A)) ) $ is nonempty. Furthermore, $F$ is an achronal topological hypersurface. Finally, following Theorem \ref{TrappedInTheAct}, $F$ is compact.  We define a complete timelike future vector field $V$ by the definition of future orientation and by rescaling to length one for a complete auxiliary Riemannian metric. Consider the map $Q: F \rightarrow S$ given by the flow of $V$. It is well-defined as $S$ is a Cauchy surface and thus achronal. Moreover, it is continuous as $V$ is smooth. The image of $Q$ lies in $\ext (A) $ by connectedness. We want to show surjectivity of $Q$ onto $\ext (A)$. First we observe that $Q$ is injective as $S$ is achronal. As $F$ and $\ext (A) $ are of the same dimension, the famous Brouwer's Theorem of invariance of domain for manifolds tells us that $Q(F)$ is open. All in all, $Q(F) $ is compact, thus closed, and open, thus all of $\ext (A)$. But this is in contradiction to $\ext (A)$ noncompact. \hfill \qed

\bigskip

\V It is worthwhile to compare Theorem \ref{VariantPenrose} to Theorem 3.2 in the interesting article \cite{EGP} by Eichmair, Galloway and Pollack, where the trapped surface in Penrose's original theorem is replaced by the assumption of existence of a MOTS, at the expense of additionally assuming the so-called 'generic condition' (whose genericity, from the initial data point of view has recently been shown in \cite{eL}, at least for a particular choice of topology on the space of initial values).

\section{Comparing different notions of asymptotic flatness}
\label{Hoop} 

\V Let us compare different notions of asymptotic flatness and resulting notions of future null infinity. It is well-known that asymptotic simplicity is too strong a requirement for many purposes. On the other hand, weak asymptotic simplicity and emptiness is too weak a condition for, e.g., obtaining global existence of solutions to semilinear partial differential equations along the lines of \cite{nGoM}. We have argued that often it can be replaced by the assumption of the existence of a conformal extension, giving up the requirement that the conformal factor equals zero on the entire boundary, to be able to treat Cauchy horizons and future null infinity on the same footing. Thus let us compare those different notions now. We call a geodesic {\em infinite} iff its affine parameter is bounded. Otherwise we call it {\em finite}. We define $LM$ to be the total space of the (nonlinear) subbundle of the null vectors in $ TM$.

\V The following two definitions are from the classical papers of Penrose, for a detailed introduction see the book of Hawking-Ellis \cite{HE}. The presentation of the following definition differs from that book, but is easily seen to be completely equivalent.

\begin{Definition}  
  
An orientable spacetime $(M,g)$ is called {\em asymptotically simple} iff it has a {\em simple extension} or {\em Penrose extension}, which in turn we define as a triple $(\psi, \hat{M}, \hat{g} )$ such that 

\begin{enumerate}
\item{$(\hat{M}, \hat{g})$ is a strongly causal spacetime and $\psi: M \rightarrow \hat{M}$ is a conformal embedding.}
\item{No maximal future- or past-directed lightlike solution $c:[0, T) \rightarrow \hat{M}$ of the geodesic equation satisfies $c([0, T)) \subset \psi (M)$; we define $\mathcal{J} := \mathcal{J} (\psi, \hat{g}) :=  {\rm exp} (D \cap L \psi (M)) \cap \partial \psi (M)  $.}
\item{Let $\Omega$ be the conformal factor for $\psi$, i.e., $\psi ^* \hat{g} = \Omega^2 \cdot g$. Then the function $ \omega := \Omega^{-1} \circ \psi^{-1} $ on $\psi (M)$ extends to a function $\hat{\omega} \in C^3 (\hat{M})$ such that $\hat{\omega} \vert_{\mathcal{J}} =0$ and $d \hat{\omega} \vert_{\mathcal{J}}  \neq 0$.}
\end{enumerate}
 
\end{Definition}

\V Note that $\mathcal{J}$ is never empty. The definition of asymptotic simplicity excludes interesting spacetimes ---  like Schwarzschild spacetime or even future causally complete spacetimes with a spacelike future infinity --- which was the reason for Penrose's second definition:

\begin{Definition}
A spacetime $(M,g)$ is called {\em weakly asymptotically simple} iff there is a simple extension $\psi: M' \rightarrow \hat{M}'$ and an open neighborhood $\hat{U}'$ of $\mathcal{J} (\psi, \hat{g}')$ such that there is an open subset $U$ of $M$ isometrical to $U':= \psi^{-1} (\hat{U}')$. 
\end{Definition}

\begin{Remark}
\label{was} 
By intersecting $U$ with the open set $I^-(\mathcal{J}, \hat{M}' )$ we can and will always assume that $U \subset I^-(\mathcal{J})$. Furthermore it is important to note that, contrary to a common misconception, the datum of $U$ is an additional one, and in general there are null geodesics of infinite affine parameter that do {\em not} end in $\mathcal{J}$. Putting an appropriate conformal factor $a$ on Kruskal spacetime yields a w.a.s. spacetime for which {\em no choice of $U$ and $\mathcal{J}$ is possible} such that all inextendible null geodesics of infinite affine parameter would end at $\mathcal{J}$:  Just choose $a=1$ outside of a small neighborhood $U$ of the left future null infinity and between $1/2$ and $3/2$ in $U$ but oscillating sufficiently wildly towards the left future null infinity to prevent any conformal extension there (e.g., by violating Penrose's peeling estimate). Kerr spacetime is an example of a  weakly asymptotically flat spacetime for which {\em no choice of $U$ and $\mathcal{J}$ is possible} such that all inextendible null geodesics would end at $\mathcal{J}$. Here the exterior region is the largest open set that can be chosen as $U$, and although there is a second open set of TIPs corresponding to the upper future null boundary in the second region of the usual diagram, no open set as above can cover it, as all null geodesics emanating from the second region have finite affine length.  
\end{Remark}

\bigskip

\V A weakly asymptotically simple spacetime is called {\em asymptotically predictable from a partial Cauchy hypersurface $S$} iff $\mathcal{J} \subset {\rm cl}( D^+ (\psi (S)))$, where $\psi$ is the Penrose extension appearing in the definition of asymptotic simplicity and thus in weak asymptotic simplicity. In Prop. 9.2.1 (p.311) of the book of Hawking-Ellis it is proven that if $S$ contains an OTS $A$, then 

\bea
\label{firstcondition}
 J^+(A) \cap \mathcal{J}^+ =\emptyset. 
\eea

\V Theorem \ref{VariantPenrose} implies that if we have a strong conformal extension $E$ and if we define $\mathcal{J}_{{\rm max}} (E)  $ as the subset of all points in $p \in \partial^+ E(M)$ such that there is a null geodesic $c$ of infinite affine length in $M$ such that $p$ is the endpoint of $E \o c$ in $\ov{E(M)} \subset N$, then   

\bea
\label{secondcondition}
J^+(A) \not\subset  \mathcal{J}^+_{{\rm max}} (E). 
\eea

\V and below we will prove that under the conditions of Theorem \ref{VariantPenrose} and the extra condition of a strong conformal extension we can show that even

\bea 
\label{thirdcondition}
J^+(A) \cap \mathcal{J}^+_{{\rm max}} (E) = \emptyset.
\eea

\V Remark \ref{was} implies that Condition \ref{thirdcondition} does {\em not} follow from the condition \ref{firstcondition}. Whereas condition \ref{firstcondition} and condition \ref{secondcondition} will not be enough to conclude the existence of black holes, the last condition is indeed sufficient.

\V As, for a conformal extension $E: M \rightarrow \hat{M}$, the target manifold $\hat{M}$ is distinguishing, two curves $c_1, c_2$ have the same endpoint $q$ in $\partial^+E(M)$ if and only if their pasts coincide. We also know that for every null geodesic $c$ in $M$, the curve $E \o c$ is a null pregeodesic in $N$ and therefore intersects $\partial^+ E(M)$ in a unique first point $q_c$. Let now $E: (M,g) \rightarrow (\hat{M},h)$ be a {\em strong} conformal extension, i.e. $h= \omega \cdot (E_* g)$ for a function $\omega \in C^\infty (E(M))$, the pushforward of the inverse of the conformal factor, such that $\a:= \sqrt{\omega}$ has a $C^k$ extension to $\hat{M}$ with causal past gradient at $\partial^+M$ (which is satisfied in particular if $E$ is strong at the point under examination). We call $q \in \partial^+ EM)$ an {\em infinity point} if $\omega$ is $C^0$ extendible by $0$ at $q$ and a {\em finiteness point} if $\omega$ is $C^0$ extendible by $a >0$ at $q$. Obviously, the set of finiteness points is open in $\partial^+ E(M)$. For well-posedness of $\mathcal{J}_{{\rm max}}(E)$ as above we recall Theorem \ref{TheEndPointRules}.

\V There is a corresponding very elementary theorem for timelike maximal geodesics, without the assumption of a conformal extension:

\begin{Theorem}
\label{wennschondennschon}
Let $c_1, c_2$ be two future timelike curves with $c_2 (0) \subset I^-(c_1) \subset I^-(c_2)$. If $c_1$ is infinite and $c_2$ is maximal, then $c_2$ is infinite as well.
\end{Theorem}

\V {\em Remark:} In particular, any two maximal curves with the same past have the same finiteness.

\V {\em Proof.} We parametrize $c_1$ by arc length and assume $ l(c_2) = C <\infty$. Choose $n >0$ such that $c_2(0) \in I^-(c_1(n))$ and a timelike curve $k$ from $c_2(0) $ to $c_1(n)$. Let $T>0$ be such that $c_1 (n+2C) \in I^-(c_2(T))$ and choose a timelike curve $K$ from $c_1 (n+2C)$ to $c_2(T) $. Then the curve $K \circ c_1 \vert_{[n, n+2C]} \circ k $ is a curve of length $>2E$ from $c_2(0)$ to $ c_2(T)$, in contradiction to maximality of $c_2$. Thus $l(c_2) = \infty$. \hfill \qed

\bigskip

\V Let us consider Kerr spacetime: Choose points $q \in \Ext ({\rm Kerr})$ and $x \in n_2 \subset \partial^+ (\BH({\rm Kerr}))$ as defined in Theorem \ref{COS}, then there is no maximal timelike geodesic from $q$ with endpoint $x$: If there were one, it would arise as a limit of maximal geodesics to $c(n)$ where $c$ is an incomplete geodesic with endpoint $x$, and those approximate $i_+$, in the limit ending there and never reaching $x$. Actually it would be contradictory to Theorem \ref{wennschondennschon} if such a geodesic curve existed, as $i_+$ is dominated by $x$. All curves ending in BH(Kerr) have finite length, as there is an isometric continuation to Block III. 
 
\begin{Theorem}
\label{SCOTness}
Let $(M,g)$ have a conformal extension $E: M \rightarrow \hat{M}$. Then a precompact set $U \subset M$ consists of visually compact points if and only if $J^-(J^+(U))$ is spatially compact. For a compact subset $R$ of $LM:= \{ v \in TM \vert g(v,v)=0 \}$, let $N_R$ be the union of images of maximal $M$-geodesics with initial values from $R$. If $J^-(c (I)) $ is spatially precompact for every maximal geodesic with $\dot{c} (0) \in R$, then $J^-(N_R)$ is spatially precompact. 
\end{Theorem}

\V {\em Proof.} We use compactness of $J^+ (\overline{U}) \cap \partial E(M)$ (by future compactness of $\overline{E(M)}$) and show that the subset $\partial_c^+ E(M)$of points $x$ on $\partial^+ E(M) $ with $ J^-(x)$ spatially compact in $E(M)$ by applying outer continuity of $I^-$ at $x$ in $\hat{M}$ to the subset $\partial C$ where $C $ is a precompact open subset of a Cauchy surface $S$ containing $J^-(x) \cap S$. The second statement follows similarly. \hfill \qed

\bigskip

\V As a last consideration, take the projection (and orbit) diagram $p: M \rightarrow D$ of any spherically symmetric asymptotically one-ended Schwarzschildian spacetime $M$ containing an OTS bounding a compact region, which can be constructed via the usual gluing procedures. We denote by $r$ the radius of the fiber and by $t$ a Cauchy temporal function on $D$ (or its lift to $M$). In a (pointed) neighborhood of $i_0$ one gets some piece of smooth future null infinity, but at some points $q$ at the conformal boundary of $D$ (which always exists), for $ R:=  e^{2u} r  $ we might get either $\lim_{p \rightarrow q} R(p) = 0$ or $\lim_{p \rightarrow q} R(p)=\infty$. However, for any open neighborhood $U$ of the closure of an initial Cauchy hypersurface we can extend $R \vert_U$ to a smooth positive function on an open neighborhood of $\overline{D}$. Of course, the corresponding spacetime $(M, \tilde{g})$ is still spherically symmetric but not TCC any more, something that can be remedied by Theorem \ref{ConformalTCCinCompact} defining a conformal factor $e^{2f}$ on $(M, \tilde{g})$, with the additional requirement that the conformal factor should increase as fast as necessary to make $t^{-1}([0;T))$ future complete, for some given $T$. The usual formulas for the second fundamental form after a conformal rescaling and the fact that the gradient of the conformal factor is past imply that the OTS continues to be an OTS. In this manner, we can construct examples of TCC spacetimes that are asymptotically Schwarzschildian and conformally extendable and contain an OTS bounding a compact region, which ensures applicability of Theorem \ref{TrappedInTheAct}.

\section{Conclusion: The difficulty of cosmic weather forecast, or: The lamentable fate of Cassandra observers}
\label{discussion}

\V There are many attempts to replace the global notions of 'mass' and 'event horizon' by quasilocal notions, i.e., those that only depend on the data in a compact subset and can be decided by a single observer (maybe even in finite time). Adopting for a moment the viewpoint of describing the dynamics of the universe in terms of classical field theories and treating observers as test particles, let us consider an analogy with the weather forecast: A notion like 'the set of places where it is going to rain at some future time' is perfectly well-defined from the purely logical point of view. However, for practical purposes we at least feel the need to complement it by a second notion that can be decided within finite time, like 'the set of places where it will rain tomorrow'. Even more, we want to have {\em predictive} criteria, i.e. methods able to detect  whether an event will arrive {\em before it arrives}. If a weather forecast informs the receivers of a storm in the very moment of its arrival such that no precautions can be taken any more, then we would call it {\em practically useless}. 

\V Singularity theorems such as Penrose's or Hawking's can also be interpreted as a censorship statement: The singularities (incomplete $C^0$-inextendible causal curves) emanating from the outer trapped surface are not visible from future null infinity. In fact, they display an even stranger form of censorship: {\em The singularity criterion itself is not visible from future null infinity}. This means, any 'Cassandra' observer able to see only the tiniest part of an outer trapped surface, consequently not agreeing on nonexistence of outer trapped surfaces assumed by other observers and trying to warn others from its effect by a minority report, is already caught in the region of finite lifetime, together with anybody who could get the warning message. Certainly, this 'brutal optimism' is some sort of cosmic censorship, but probably a more cruel one than usually expected. And it shows that the pragmatic power of predictivity of the Penrose singularity theorem for Einstein-Maxwell theory in the asymptotically flat case is very limited, actually its statement is practically useless in the sense of the weather forecast criterion above.

\bigskip
\bigskip

\begin{center} 
{\Small
Olaf M\"uller, Institut f\"ur Mathematik, Humboldt-Universit\"at zu Berlin, D-10099 Berlin \\ E-mail: mullerol@math.hu-berlin.de
}
\end{center}

\end{document}